# Low-Power Temperature Control by Chiral Ferroelectric Nematic Liquid Crystal Windows

Md Sakhawat Hossain Himel [1], Rohan Dharmarathna[1], Netra Prasad Dhakal[1], Kelum Perera[2], S. Sprunt[2], J.T. Gleeson[2], Robert J. Twieg[3], Antal Jákli[1,2, *]

[1]Materials Sciences Graduate Program and Advanced Materials and Liquid Crystal Institute, Kent State University, Kent, Ohio 44242, USA

[2]Department of Physics, Kent State University, Kent, Ohio 44242, USA

[3]Department of Chemistry and Biochemistry, Kent State University, Kent, Ohio 44242, USA

*: Author for correspondence: ajakli@kent.edu

## Abstract

Low power consumption is critical for smart windows for temperature control and privacy. The recently discovered ferroelectric nematic ($N_F$) liquid crystals exhibit strong coupling of the ferroelectric polarization with electric fields, making them promising candidates for energy-efficient electrochromic devices.

Here we investigate the electrochromic properties of a room temperature chiral ferroelectric nematic ($N_F^*$) liquid crystal in films with in-plane electrodes, where the electric field is perpendicular to the helical axis. We demonstrate that smart windows based on this material can regulate interior temperatures within a 10 °C range using only 50 mW/m² specific power, achieving 50% larger temperature modulation and 50–100 times lower power consumption than polymer dispersed and polymer stabilized liquid crystal windows. These findings suggest that chiral $N_F$ liquid crystals offer a highly efficient approach for smart window applications, potentially surpassing existing electrochromic technologies in energy efficiency and thermal regulation.

# 1. Introduction

Traditional windows in buildings allow for natural light to enter but provide insulation against heat transfer across the building envelope. Smart windows can enhance both these attributes by providing dynamic control of light transmission and improved thermal performance[1–3]. They may also increase both energy and cost efficiency[3,4] by reducing heat gain inside buildings and reflecting unwanted radiation[5,6]. There are three main classifications of smart windows: electrochromic, thermochromic or photochromic.

Electrochromic windows employ electrical signals to adjust their color and/or opacity[7–19]. Among them reversible metal electrodeposition (RME) devices have low specific power such as $< 1.4 Wm^{-2}$ for Ag-based RMEs.[20] Thermochromic materials[21–29] have optical properties that vary with their temperature. For example, vanadium dioxide ($VO_2$) based thermochromics rely on a reversible, metal to insulator transition that alters optical transmittance in the near infrared (NIR) band[30,31]. Photochromic smart windows use organic dyes or metal oxide particles that alter the absorption characteristics under varying radiation conditions[32,33].

Among electrochromic materials, liquid crystal (LC) based systems have gained considerable attention in the last decades due to their fast responsiveness with only low electrical power requirements.[34] To date, three electrochromic LC technologies have been employed: polymer dispersed liquid crystal, (PDLC)[35–38], polymer stabilized liquid crystal (PSLC)[39–42], gel-glass dispersed (GDLC)[43] and dye-doped liquid crystals[44–50]. PDLCs and PSLCs switch between opaque and transparent states, whereas dye-doped LCs regulate light absorption without obscuring the view as of opaque states.

In PDLCs micrometer size LC droplets are dispersed within a continuous, isotropic polymer matrix. The polymer to liquid crystal volume ratio is ~1:1[51]. In the absence of an applied field, the average molecular direction (so-called director) of the individual LC droplets is random. This results in an inhomogeneous refractive index of the PDLC, which strongly scatters light and makes the PDLC film opaque. Under a sufficiently high applied voltage (about 30-60V for a 20-30μm thick PDLC films), the director becomes aligned by the electric field [52] as illustrated in Figure 1(a). If the refractive index of the aligned LC and of the polymer matrix match, the PDLC medium becomes transparent.

PSLCs contain a small volume fraction (usually less than 10%) of polymer that forms a network within a continuous liquid crystal (LC) matrix. Depending on the phase and the alignment of the liquid crystal molecules when polymerization takes place, PSLCs can switch from opaque to transparent (normal mode), or from transparent to opaque (reverse mode) when a high enough voltage (typically 10 - 40V) [53,54] is applied, as shown in Figure 1(b). In both PSLCs and PDLCs the structure depends on phase separation; therefore, they are prone to long-term stability issues. Furthermore, the opaque states obscure the view, so they are suitable for privacy windows but not for exterior windows for cars or houses.

The normally dark mode dye-doped nematic liquid crystals change their tint by switching the director and the orientation of the incorporated dichroic dye molecules with an applied field from the original planar orientation (parallel to the substrates) to homeotropic (perpendicular to the substrates). In the planar orientation the dye molecules absorb light, whereas in the homeotropic state they transmit light with minimal absorption (see Figure 1(c)). The advantage of this switching mechanism is that both the dark and bright states provide an unobscured view, so they are suitable for smart exterior window applications. In fact, solar powered and automatically adjusting adaptive films for glazing applications have previously been proposed[44]. Their disadvantages include the light absorption-induced heating that decreases the cooling efficiency. Based on this heating effect, a smart reflector[55] consisting of a chiral nematic (cholesteric) liquid crystal with temperature sensitive helix pitch was proposed by Meyer et al[55]. Light entering the liquid crystal is absorbed by the dye, generating heat which raises the temperature of the liquid crystal. The resulting change in the helix pitch of the cholesteric causes an increase in its reflectivity, reducing the intensity of light that can be absorbed by the dye. This negative feedback stabilizes the light intensity transmitted into an interior separated by the smart window. Although such "smart reflector" would keep the light intensity constant in a room, it would not control the temperature, which is desired for most cases.

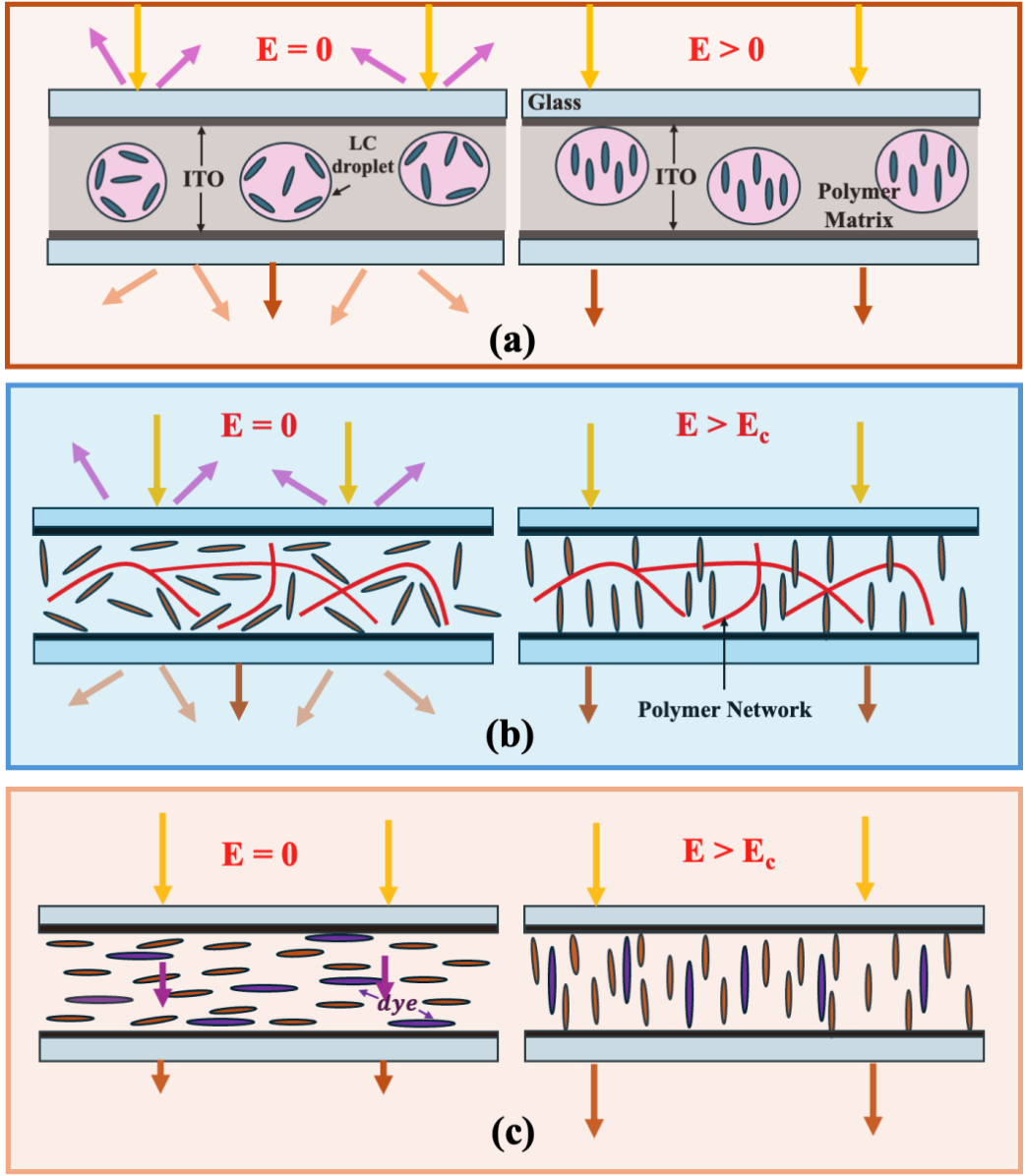

*Figure 1: Schematic illustration of the structure and optical properties of polymer dispersed liquid crystals (PDLCs) (a), polymer stabilized liquid crystals (PSLCs) (b), dye-doped liquid crystals (c) in OFF state (left) and under AC electric fields applied between indium tin oxide (ITO) electrodes coated on glass substrates (right). Orange, pink, magenta, brown and purple arrows indicate the incoming, back scattered, forward scattered, transmitted and absorbed light, respectively.*

In this paper we propose a highly efficient, liquid crystal window that can be used to keep the internal temperature of a room or car low at high external temperature but would let the light enter in cold weather. It is based on electrically controlled reflection of light as opposed to light scattering or absorption. This technology exploits the recently discovered ferroelectric nematic ($N_F$) liquid crystals [56–59] that are extremely sensitive to electric fields and have already showed promise for high-speed electro-optical modulators[60], microlenses[61,62] and photovoltaic[63] devices. Chiral ferroelectric nematic ($N_F^*$) liquid crystals form a helix with pitch that depends sensitively on temperature, as well as the concentration and twisting power of chiral doping agent. $N_F^*$ materials can exhibit wavelength variable reflectivity spanning the visible spectrum. Furthermore, this reflectivity is reversibly tunable over $\Delta\lambda \sim 150$ nm wavelength range with only a small electric field applied perpendicular to the helical axis[64,65]. These and other results on $N_F^*$ materials[66] can be understood within the context of electrically tunable, multiple photonic band-gaps[65]. Recently we demonstrated electrical tuning reflectivity of room temperature $N_F^*$ mixtures, thus offering practical applications[67]. With these formulations, depending on the frequency of the applied field and on the mixture composition, one can either tune the wavelength or the amplitude of the reflection or both.

In this study we demonstrate that the reflectivity can be doubled by stacking $N_F^*$ electrochromic films with left and righ handed helices and investigate how in-plane electric field controlled reflectivity can be utilized for temperature control of an interior separated by $N_F^*$ windows. For these studies we chose a custom-made room temperature $N_F$ mixture, designated KPA-02, doped with 3.2% chiral dopants R-5011 or S-5011 from Merck. KPA-02 mixed with chiral dopants was previously shown to tune both the wavelength and the intensity of the selective reflection under in-plane DC field, or only the reflectivity without wavelength changes under 100 Hz AC fields [67] as illustrated schematically in Figure 2. For control purposes we also studied standard PDLC and PSLC films. We show $\leq 50 \text{ mW/m}^2 \, AC$ power consumption and a temperature tuning range larger than that of both PSLC and PDLC windows. This substantially improved performance yields significant promise for a wide variety of applications, most notably next-generation architectural windows.

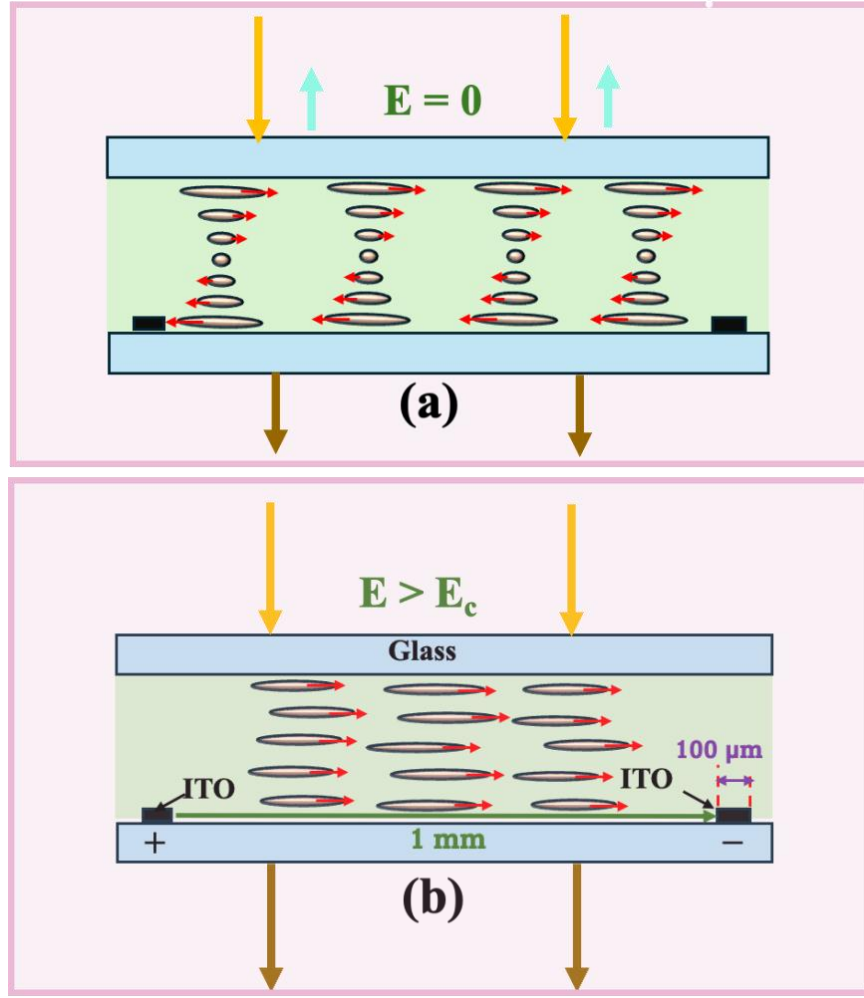

*Figure 2: Schematic illustration of the structure and optical properties of $N_F^*$ materials in the in-plane geometry at E=0 (a) and E>E$_c$ (b). Red arrows show the direction of the spontaneous polarization. Green arrow shows the electric field in half period when it is from left to right. Orange, brown and light blue arrows indicate the incoming, transmitted and reflected light, respectively. The molecular structure depicted is expected to be true inside the film, far away from the substrates.*

## 2. Results and Discussion

The composition of KPA-02 and of the control PDLC and PSLC materials, and the corresponding cell preparations, are described in the Materials and Methods section.

Results on wavelength dependence of the reflectivity of KPA-02 + 3.2% R5011 (top inset), KPA-02 + 3.2% S5011 (bottom inset) and both samples stacked vertically (main pane) are shown in Figure *3*. All samples exhibited maximum reflectivity in the wavelength range of 430-440 nm.

The maximum reflectivity with no electric field is ~33% for the individual R and S doped samples, and ~70% when both R and S doped samples are stacked vertically. For all three cases, the maximum reflectivity values decrease with the application of 100 Hz sinusoidal in-plane electric field and the reflectivity drops to just a few percent at 0.14 V/μm. During this reflectivity variation, the wavelength for maximum reflectivity remained unchanged, i.e., the color reflected remained the same.

When the field is decreased below the level that completely unwinds the helix, we observe a transient scattering state, and the equilibrium color (OFF-state) recovers only after several hours. However, if the field remains below the unwinding level, when switched off, the time to recover the original, zero-field reflectivity is less than one minute. When the field changes between non-zero and less than unwinding level, the reflectivity change is complete within one second, i.e., the tuning is rapidly reversible.

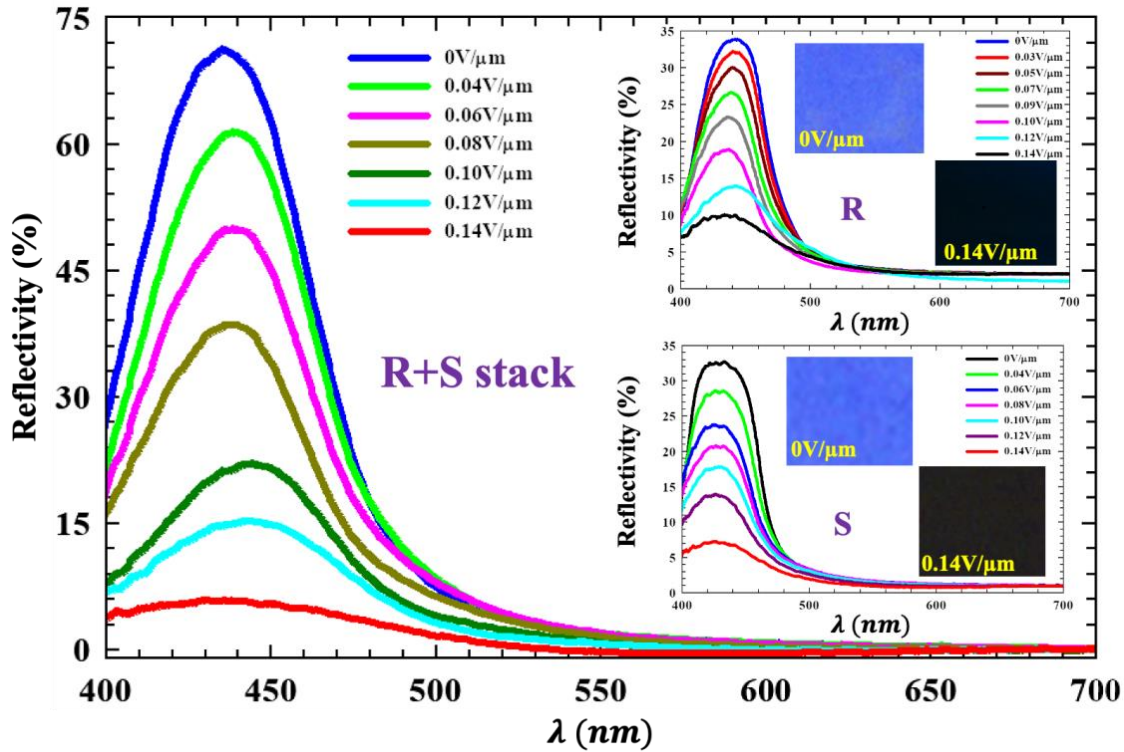

*Figure 3: Wavelength dependence of the reflectivity of KPA-02 + 3.2% R5011 (top inset), KPA-02 + 3.2% S5011 (bottom inset) and both samples stacked vertically (main pane). Small insets: microscopy images (white light illumination: size: ∼ 120 μm × 100 μm) showing colors reflected at zero and 0.14 V/μm, 100 Hz electric fields.*

Insets of Figure *3* represent ~ 120 µm × 100 µm area reflection microscopy images of R and S-doped samples under white light illumination. In both cases the images are bright blue at zero field, and dark at 0.14 V/µm 100 Hz in-plane field. These results are consistent with previous results[67] obtained with KPA-02 and 3.2% BDH1281, indicating that the helical twisting power of BDH1281, R5011 and S5011 are similar under these circumstances. [67]The decrease of the reflectivity is the result of the field-induced distortion of the helical structure but without change of its pitch. [67]

The effect of the tuning of the reflectivity on the temperature control of a room separated from external light by an $N_F^*$ window with right-handed helix ($N_F(R)$) and a stack with right and left-handed helices ($N_F(R+S)$) is demonstrated in Figure *4* in comparison with PDLC and PSLC windows. To mimic a room with a single window, a homemade thermally insulated box with a hole in the front was covered by one of the LC windows. We monitored the internal temperature of the box as a function of voltage applied on the LC windows, while the box was illuminated by a white light with 40 $mW/cm^2$ irradiance corresponding to an average Sunlight intensity in Ohio. The set-up used to simulate the temperature control of $N_F^*$, PSLC and PDLC based systems are shown in Figure 6(b) in the Materials and Methods section. The box temperature as a function of specific power determined as $P/A = \frac{V^2}{R \cdot A}$, where $V$ is the applied voltage, $R$ is the sample resistance and $A$ is the area of the window, is plotted in Figure *4*.

In all studied windows (the R-, S-, and R+S doped KPA-02 cells with in-plane electrodes, and the control PSLC and PDLC cells) the liquid crystal films had 30 µm thicknesses and the temperature of the exterior was kept at 22 °C before illuminating the window with a Xenon gas lamp light. In each case, the zero field states are the least transparent due to the initial reflectivity of the $N_F^*$ samples and the initial opacity of the control PSLC and PDLC films. At zero field under light illumination the equilibrium internal temperatures were 25 °C for $N_F(R+S)$, 26.5 °C for the $N_F(R)$ windows, and 28 °C for the control PSLC and PDLC films. This indicates that the reflection of the $N_F^*$ cells in only about 100 nm wavelength range are more effective to decrease the heating effect than of the scattering of the PDLC and PSLC films. This is likely due to the back-scattering of the PDLC and PSLC samples. The 1.5 °C lower temperature achieved by the $N_F(R+S)$ stack is due to the larger reflectivity. The temperatures in the clear states are 32.5 °C for $N_F(R)$, 33.5 °C for the PSLC and 36 °C for the $N_F(R+S)$ and PDLC samples. This indicates that the PDLC sample has the clearest state. The differences of the temperatures at the clear-state of the $N_F$(R+S) and the $N_F$(R) samples, that are equally clear, can be attributed to the additional glass plates that reflect the light scattered from the inside of the box, thus trapping more heat in the box. Comparing the temperature ranges ($\Delta T$) can be tuned electrically in the different samples, we see that it is the largest ($\Delta T \approx 10\ °C$) for $N_F$(R+S) and smallest ($\Delta T \approx 5.5\ °C$) for the PSLC. The most important observation is that the highest specific powers to achive the clear states are only $50\ mW/m^2$ for the $N_F^*$ samples, while they are $2\ W/m^2$ and $10\ W/m^2$ for the PDLC and PSLC samples, i.e., they are 40 and 200 times larger, respectively. Here we note that the power consumption of the PDLC and PSLC samples we measured agree with prior measurements that reported the specific power of PDLCs in the range of $5 - 20 \frac{W}{m^2}$[68–71] and $2 - 4 \frac{W}{m^2}$[72,73], respectively.

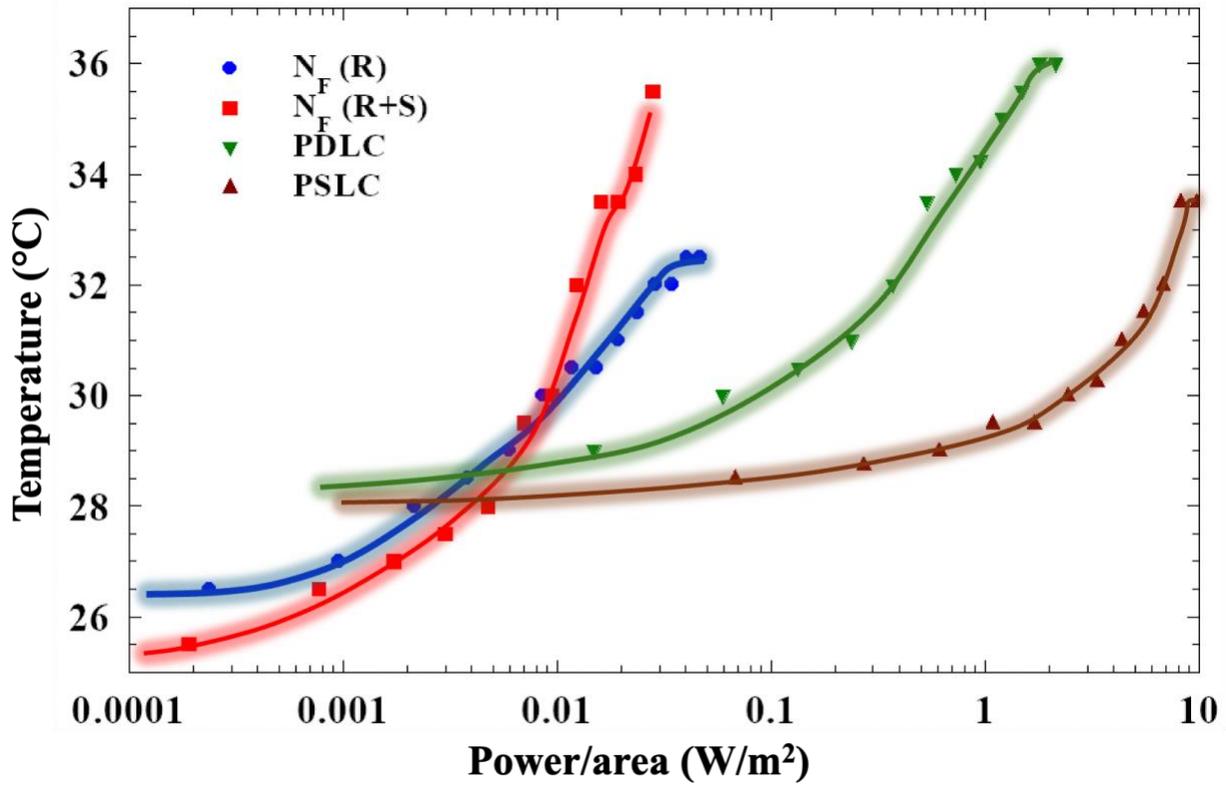

Figure 4: *Internal temperature of a thermally isolated box as a function of power/area $P/A = \frac{V^2}{R \cdot A}$, where V is the voltage applied on various $N_F^*$, PSLC and PDLC windows, R is the LC resistance and A is the area of the LC window, while the box was illuminated by a white light with $40\ mW/cm^2$ irradiance.*

To the best of our knowledge similarly low power consumption was observed only in dye doped [44] and plasmonic nanoplate dispersed[74] nematic liquid crystals. Although in those windows the effect of the electric field on the temperature regulation was not reported, in the nematic plasmonic nanocolloids it was estimated that 9% of a standard building's energy can be saved.

The remarkable low power consumption in our $N_F^*$ windows may be related to the high sensitivity of the helical structure of the helix to the electric field applied perpendicular to the helix and to the in-plane field geometry. In sandwich cells (out-of plane field) such as used in the PDLC and PSLC films, the resistance $R_s = \varrho \cdot \frac{d}{A}$, where $\varrho$ is the resistivity of the material and $d \approx 30$ μm is the liquid crystal film thickness. For $N_F^*$ cells with in-plane electrodes, the resistance is $R_{IP} = \varrho \cdot \frac{L}{d \cdot w} = \varrho \cdot \frac{L^2}{d \cdot A}$, where $L = 1\ mm$ is the distance between the in-plane electrodes, $w$ is width of the cell, so the area normal to the light is $A = L \cdot w$. This means that for sandwich cells $\frac{P}{A} = \frac{V_s^2}{\varrho \cdot d}$, while for in-plane cells $\frac{P}{A} = \frac{V_{IP}^2 d}{\varrho L^2}$. Comparing $N_F^*(R)$ with the PDLC of the same film thickness we get that the ratio of the power per unit area is $\frac{P_s}{P_{IP}} = \frac{V_s^2 \varrho_{N_F^*(R)}}{V_{IP}^2 \varrho_s} \cdot \frac{L^2}{d^2}$. With the known $L, d, V_S$ and $V_{IP}$ and the measured $P_S$ $P_{IP}$ values we can also calculate the ratios of the resistivity values as $\frac{\varrho_{N_F^*(R)}}{\varrho_S} = \frac{V_{IP}^2}{V_S^2} \cdot \frac{d^2}{L^2} \cdot \frac{P_S}{P_{IP}}$. Taking into account that in the PDLC and PSLC sandwich cell the maximum applied voltage is $V_S = 60V$, while in the in-plane $N_F^*$ cells $V_S = 140V$, furthermore from the measurement results that $\frac{P_{PDLC}}{P_{N_F^*}} \approx 40$ and $\frac{P_{PDLC}}{P_{N_F^*}} \approx 200$, we get that $\varrho_{N_F^*(R)} \approx 0.2 \varrho_{PDLC}$ and $\varrho_{N_F^*(R)} \approx \varrho_{PDLC}$. This means that the required power consumption is much smaller in the $N_F^*$ cells even though their resistivity is 5-times smaller than of a PDLC. Therefore, it is possible that as more and better $N_F^*$ formulations with higher resistivity become available, their power consumption will decrease substantially rendering them even more competitive than the presently available electrochromic windows.[20,34,75,76]

We note that normal mode PDLC and PSLC windows are opaque in OFF state and the outside view will be obscured. Recently polymer stabilized cholesteric liquid crystal windows with interdigitated electrodes that can dynamically transmit both visible light and infrared light in OFF state have been proposed. In those windows the specific power to reach opaque focal-conic state was $2.3\ W/m^2$, i.e., about 50 times larger than our $N_F^*$ windows. [72]

Finally we note that although the presented $N_F^*$ windows with $1\ mm$ gap in-plane electrodes require larger voltage than the PDLC and PSLC sandwich cells, interdigitated electrodes with $0.1\ mm$ gap and $\leq 10\ \mu m$ wide electrode strips will require only $15\ V$ while the small ITO electrodes will not be visible with naked eye.

## 3. Conclusions

In this work we have studied the electro-optic properties of a room temperature, chiral, ferroelectric nematic liquid crystal under in-plane electric fields applied perpendicular to the helix. We further explored how these properties can be used to craft smart windows to assist in regulating the temperature of an interior space separated from the exterior environment by the smart window. We find that the $N_F^*$ window that requires using only 50 mW/m² electric power per area, leads to superior temperature modulation. In our experiments we maintained 3 °C lower temperatures at zero power consumption and 50% larger temperature tuning with 50-100 times less power consumption than that of polymer dispersed, and polymer stabilized liquid crystal windows. [49,44]This novel technology also favorably compares to electrochromic windows, thus highlighting its promise for new, efficient. smart window applications.

We expect that a stack of several $N_F^*$ formulations with larger helical pitches should reflect light in wider wavelength range and are expected to further increase their heat controlling efficiency.

## 4. Materials and Methods

KPA-02 contains 60 wt% single component ferroelectric nematic liquid crystals (FNLC) compound RT12155 [4-[(2,6-difluoro-4-cyanophenoxy) carbonyl] phenyl 2-n-propoxybenzoate][67] and 40 wt% of a commercially available nematic LC HTG-135200-100 (clearing point $T_c = 97°C$) purchased from HCCH. On cooling from the isotropic phase, KPA-02 transitions to N$_F$ phase at 47°C and remains stable at room temperature. The spontaneous polarization of KPA-02 is measured to be $P_s \approx 0.044 \frac{C}{m^2}$ at 33°C. [67] To induce chirality, $c \approx$ 3.2 $wt\%$ commercially available right - and left - handed chiral dopants R-5011 and S-5011 were added to KPA-02. The helical twisting powers of these dopants on KPA-02 is $HTP \sim 120 \mu m^{-1}$, therefore the pitch of the chiral materials is $p = (c \cdot HTP)^{-1} \approx (0.032 \cdot 120)^{-1} \approx 0.26$ μm. The molecular structures of the host FNLC and the chiral dopants are shown in Figure 5(a).

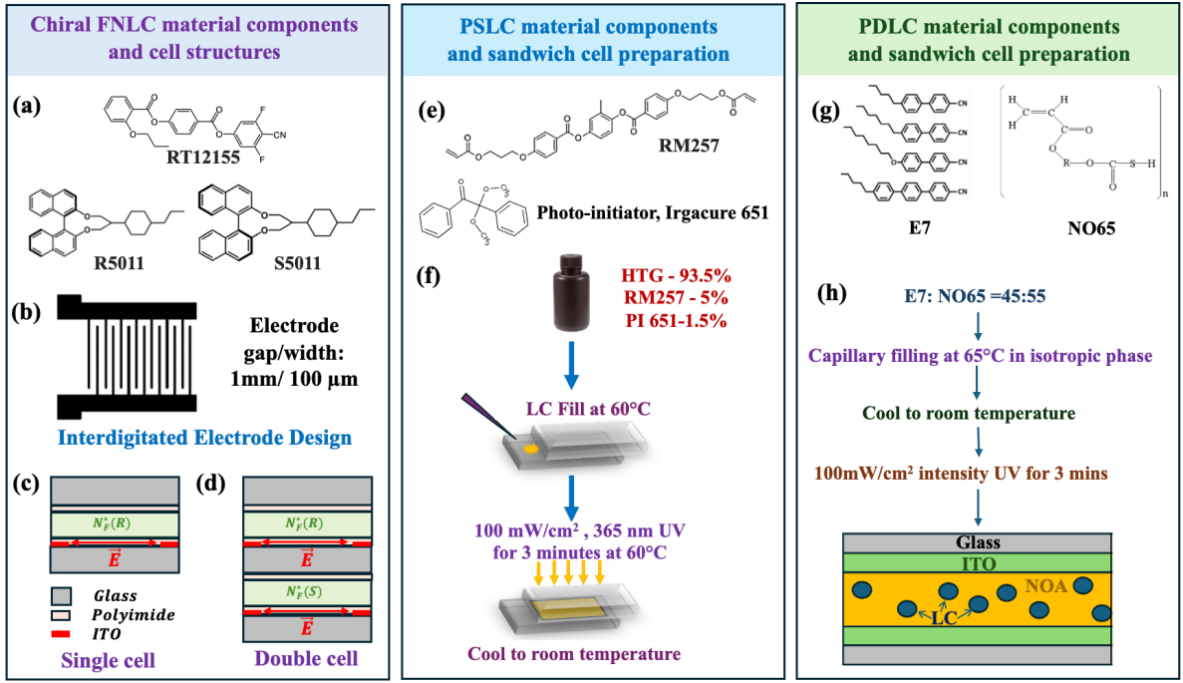

*Figure 5: Molecular structures of the materials and illustrations of the LC cell preparations. (a) Molecular structures of the host $N_F$ material RT12155, right-handed chiral dopant R-5011, and left-handed chiral dopant S-5011. (b) Preparation of the photomask with interdigitated electrodes. (c) Cross sectional view of a single 30 μm thick $N_F^*$ (R) cell with in-plane electrodes. (d) Cross sectional view of a stack of $N_F^*$ (R) and $N_F^*$ (S) cells with out-of-plane electrodes. (e) Molecular structure of RM257 and photo-initiator Irgacure 651. (f) Schematics of the preparation of normal mode PSLC cells. (g) Molecular structures of the components of E7 and NO65. (h) Diagram showing the preparation of PDLC cell with the schematic presentation of a PDLC structure.*

The liquid crystals were filled in between two polyimide (PI 2555) coated glass substrates separated by 30μm spacers. There were single layer of $N_F^*$(R) and $N_F^*$(S) cells and a stack of $N_F^*$ (R) and $N_F^*$ (S) films, $N_F^*$ (R + S) cell, with in-plane interdigitated electrodes (see Figure 5(c,d)). One substrate had no electrode, whereas the other one had 100 μm wide interdigitated in-plane indium tin oxide (ITO) electrodes separated by $1mm$ gaps, as shown in Figure 5(b). The interdigitated electrodes (IDE) were designed by AutoCAD and prepared in cleanroom by standard photolithography technique. The PI 2555 coatings were applied by spin coating. The spin-coated substrates were soft baked at 90 °C for 5 minutes, hard baked at 275 °C for 1 hour, then rubbed unidirectionally using a block covered with velvet cloth to align the liquid crystal parallel to the substrates and the rubbing. The cells were capillary filled on heat bench by the $N_F^*$ mixtures.

For the control polymer stabilized liquid crystal (PSLC) mixture we used the same commercially available nematic mixture HTG-135200-100 that in 40% was used in KPA-02 as well. We added 5 wt% photo-polymerizable LC monomer RM 257 (2-Methyl-1,4-phenylene bis(4-(3-(acryloyloxy)propoxy)benzoate)) purchased from Synthon Chemicals, and 1.5 wt% of photo-initiator Irgacure 651 (2,2-dimethoxy-2-phenylacetophenone) from Sigma-Aldrich. The chemical structures of RM257 and Irgacure 651 are shown in Figure 5(e). For the construction of PSLC sandwich cell, two ITO coated glass substrates with parallel rubbed PI2555 coating separated by 30μm spacers were used. The PSLC mixture was capillary filled in the isotropic phase at 60°C. The cell was then photopolymerized under 365 nm, 110 $mW/cm^2$ intensity UV light (Blak-ray, Model B-100AP) for 3mins at 60 °C as shown in Figure 5(f).

The polymer dispersed liquid crystal (PDLC) control sample is composed of 45 wt% of nematic LC E7 purchased from Merck, and 55 wt% Norland optical adhesive (NO 65) from Edmund Optics. The molecular structures are presented in Figure 5(g). The mixture was first capillary filled into 30 μm sandwich cell in isotropic phase at 65°C, then it was cooled to room temperature and UV cured 100 mW/cm² for 3 minutes. The schematic of the cell fabrication process and the cross-section of various segments of the cell with PDLC composition are shown in Figure 5(h).

Figure 6 shows the experimental setup to measure the voltage dependent reflectance of the $N_F^*$ cells. An HP 33120A function generator and an FLC F20AD voltage amplifier were used to apply the electric fields. The reflectivity *R(λ)* was obtained using an OceanOptics VIS-IR spectrophotometer.

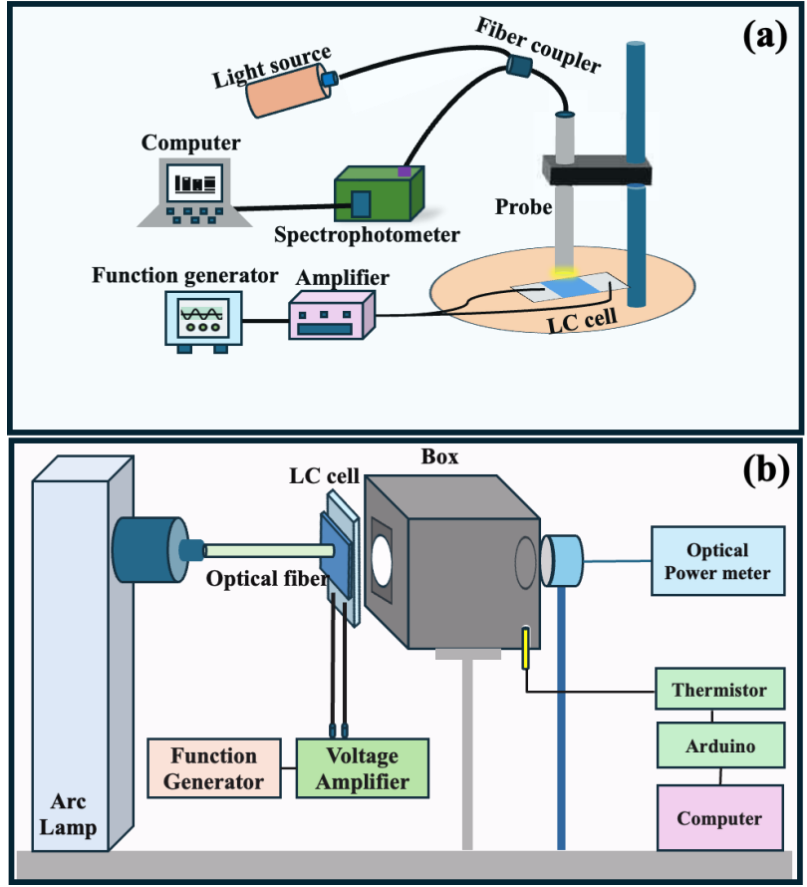

*Figure 6: Experimental setups. (a) Measurement of the voltage dependent reflectivity of the chiral FNLC cells. (b) Measurement of voltage dependent temperature variations within a box mimicking a room with smart windows.*

The schematics of the set-up to measure the temperature control efficiencies of the $N_F^*$, PSLC and PDLC based systems is shown in Figure 6(b). It is comprised of an Oriel Arc Lamp Housing (Model 66021) equipped with an Ushio UXL-451-O 450W Xenon gas lamp, and powered by the Oriel Universal Power Supply (Model 68820) to generate a maximum flux of 13000 lm. A homemade thermally insulated box with a hole in the front covered by one of the LC windows mimics a room with a single window. A hole in the back open to the power meter measures the transmitted optical power. A thermistor inserted in the lower back records the temperature within the box. The output data is processed by an Arduino microcontroller and a computer. Temperature values were measured fifteen minutes after changing the applied voltage to allow for equilibration. The resistances of the LC cell were measured with a Keithly 6517B electrometer.

## 5. Acknowledgement

This work was supported by US National Science Foundation grant DMR-2210083.